\documentclass[aps,prl,reprint, amsmath, amssymb,superscriptaddress]{revtex4-2}
\usepackage{xr-hyper}

\usepackage{graphicx,tikz,placeins}
\usepackage{mathrsfs}
\usepackage[english]{babel}
\usepackage{amsmath,amssymb,amsfonts}
\usepackage{color}
\usepackage{lipsum}
\usepackage[colorlinks=true,linkcolor=red,urlcolor=blue,citecolor=cyan,pdfusetitle]{hyperref}

\DeclareMathOperator{\var}{Var}
\DeclareMathOperator{\covar}{Cov}

\makeatletter
\newcommand*{\addFileDependency}[1]{
\typeout{(#1)}
\@addtofilelist{#1}
\IfFileExists{#1}{}{\typeout{No file #1.}}
}\makeatother

\begin{document}

\title{Pareto-optimal control strategies in intrinsically nonequilibrium systems}

\author{Gustavo A. L. For\~ao}
\email[]{gforao@usp.br}
\affiliation{Universidade de São Paulo,
Instituto de Física,
Rua do Matão, 1371, 05508-090
São Paulo, SP, Brazil}

\author{Carlos E. Fiore}
\affiliation{Universidade de São Paulo,
Instituto de Física,
Rua do Matão, 1371, 05508-090
São Paulo, SP, Brazil}

\author{Jonas Berx}
\affiliation{Niels Bohr International Academy, Niels Bohr Institute,
University of Copenhagen, Blegdamsvej 17, 2100 Copenhagen, Denmark}

\date{\today}

\begin{abstract}
Thermodynamic control is typically formulated as the optimisation of a single objective, yet competing costs rarely admit a common optimum, so single-objective control captures only one corner of the achievable performance space. We develop a general framework for multi-objective thermodynamic control of intrinsically nonequilibrium systems that maps out the full Pareto front of optimal control strategies. We show that Pareto-optimal protocols generically consist of smooth branches connected by boundary jumps, and that the relative weights of the objectives combine with the physical parameters into a single intrinsic scale that alone governs the trade-off. Remarkably, this scale plays a double role: it parametrizes a single functional form that generates the entire front, and it defines control equivalence classes, in which systems with different parameters but the same scale share identical optimal strategies. We illustrate the framework for two paradigmatic systems that are experimentally accessible: transport of an active particle in a harmonic trap, and a cyclic quantum-dot engine. For both, we obtain the Pareto front and optimal strategies in closed form.
\end{abstract}

\maketitle

Far from thermodynamic equilibrium -- where many physical, chemical, and biological processes naturally operate -- the ability to precisely control the dynamics of a system is crucial, e.g., for the fundamental understanding of nonequilibrium behavior or for the ability to enable engineered applications, ranging from energy-efficient nanoscopic machines to the design of exotic materials or artificial biosystems. Thermodynamic control theory therefore aims to design driving protocols that optimise a chosen objective: minimising work input or fluctuations, or maximising power and/or efficiency. While it is typically the control action itself that drives a system out of equilibrium~\cite{Schmiedl2007,Gomez2008,Solon2018}, controlling \emph{intrinsically} nonequilibrium systems has only recently been the subject of investigation~\cite{Olsen2025,Loos2025,Loos2025_2}.

At the mesoscale, observables such as power, fluctuations, and efficiency are constrained by trade-offs such as thermodynamic and kinetic uncertainty relations~\cite{Barato2015,DiTerlizzi_2019,Vo_2022}, and cannot be simultaneously optimised. In general, however, such relations do not fully capture the trade-off structure of \emph{driven} systems, and emergent trade-offs (Pareto fronts) are not expected to be captured by such uncertainty relations~\cite{Solon2018}. More broadly, objectives are typically system-dependent and mutually incompatible: improving one often degrades another, so single-objective optimisation captures only a partial view of the control landscape. Although such competing objectives are legion out of equilibrium~\cite{Proesmans2023,Berx2024}, the trade-offs they generate remain largely unexplored in thermodynamic optimal control.

\begin{figure}[t!]
    \centering
    \includegraphics[width=\linewidth]{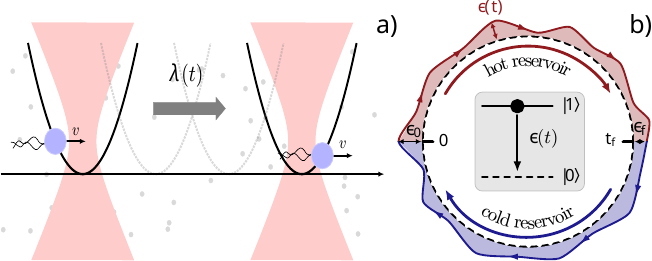}
    \caption{Controlled nonequilibrium systems. {\bf(a)} Active particle (blue) dragged through a fluid by a harmonic trap induced by optical tweezers (red) according to a protocol $\lambda(t)$. {\bf(b)} A two-level quantum dot (gray box) in alternating contact with hot (red) and cold (blue) reservoirs; the level spacing $\epsilon(t)$ between the ground state (dashed lines) and excited state can be tuned cyclically (clockwise) according to the coloured arrows; the instantaneous spacing $\epsilon(t)$ is illustrated by the wavy outer borders.}
    \label{fig:cartoons}
\end{figure}

In this Letter, we formulate multi-objective thermodynamic control as a variational problem of an effective cost functional whose solutions trace the entire trade-off. This reveals a structural feature invisible to single-objective optimization: the weight assigned to each competing objective enters the optimal protocol only indirectly, being absorbed into effective physical parameters of the system, which combine into a single intrinsic scale. This scale governs the trade-off and labels equivalence classes of control problems---physically distinct systems collapsing onto the same optimal protocol. We demonstrate this in two paradigmatic systems, shown in Fig.~\ref{fig:cartoons}:  an active particle dragged through a fluid by an optical tweezer, and a cyclic quantum-dot heat engine.


\emph{General setup---}
To establish our framework for multi-objective control, we introduce a control field $\boldsymbol{\mathcal{U}}(\mathbf{r},t)$ representing all tunable parameters of the system. Performance is then characterised by $N$ competing cost functionals $\{\mathcal{C}_i[\boldsymbol{\mathcal{U}}]\}_{i=1}^{N}$, representing, e.g., mean work, dissipation or fluctuations. The set of optimal trade-offs between these objectives, i.e., the Pareto front, is found through minimisation of the scalarisation of the costs, $\Omega[\boldsymbol{\mathcal{U}}] = \sum_{i=1}^{N} \beta_i \mathcal{C}_i[\boldsymbol{\mathcal{U}}]$, where $\beta_i\ge0$ with $\sum_i\beta_i=1$ determine the relative weight of each thermodynamic objective. The control field is typically constrained by physical requirements such as fixed initial and final states or dynamical constraints, encoded through $M$ constraint functionals $\{\mathcal{V}_j[\boldsymbol{\mathcal{U}}]=0\}_{j=1}^{M}$. These constraints are enforced by introducing the augmented functional $\tilde{\Omega}[\boldsymbol{\mathcal{U}}]=\sum_{i=1}^{N} \beta_i \mathcal{C}_i[\boldsymbol{\mathcal{U}}]
-\sum_{j=1}^{M} \mu_j \mathcal{V}_j[\boldsymbol{\mathcal{U}}]$, where $\mu_j$ are Lagrange multipliers. Pareto-optimal protocols then correspond to stationary points satisfying $\delta\tilde{\Omega}=0$. Taking the functional derivative with respect to the control field $\boldsymbol{\mathcal{U}}$ yields the general optimality condition
\begin{equation}
\sum_{i=1}^{N}\beta_i
\frac{\delta\mathcal{C}_i}{\delta\boldsymbol{\mathcal{U}}(\mathbf{r},t)}
=
\sum_{j=1}^{M}\mu_j
\frac{\delta\mathcal{V}_j}{\delta\boldsymbol{\mathcal{U}}(\mathbf{r},t)}.
\label{eq:functional_stationarity}
\end{equation}

Eq.~\eqref{eq:functional_stationarity} is the starting point of the analysis, and three consequences are worth highlighting. First, when the costs admit a local representation $\mathcal{C}_i=\int \mathcal{L}_i\,dt$, the functional derivative reduces to the Euler--Lagrange equation of the effective Lagrangian $\mathcal{L}_{\rm eff}=\sum_i\beta_i\mathcal{L}_i$. Second, discontinuous jumps in the optimal protocol arise from two mathematically distinct mechanisms: for memoryless cost functionals, jumps originate from the constraint terms on the r.h.s. of Eq.~\eqref{eq:functional_stationarity}, where the boundary-supported $\delta\mathcal{V}_j/\delta\boldsymbol{\mathcal{U}}$ generate Dirac-delta contributions enforcing the endpoint data; for cost functionals with finite-memory kernels, jumps additionally emerge from the left-hand side as singular contributions required to invert the smoothing kernel on a finite interval. Third, the optimal protocol depends on the system parameters only through the ratios of cost coefficients, thereby organizing distinct control problems into equivalence classes — each class spanning the Pareto front through a single protocol form (see SM for a detailed derivation).

We now illustrate this framework by considering two paradigmatic examples that can be solved exactly: dragging an active particle with a harmonic trap and a thermodynamic machine realised by cyclically modulating the energy level of a quantum dot coupled to a metallic lead at different temperatures.

\emph{Active particle model---}We set the stage by considering a self-propelled particle with velocity $v(t)$ trapped by a harmonic potential $V(x,\lambda) = k(x-\lambda)^2/2$ with stiffness $k$, centered at position $\lambda$, as sketched in Fig.~\ref{fig:cartoons}(a).

The dynamics of the particle is described by the one-dimensional overdamped Langevin equation:
\begin{equation}
    \label{eq:langevin_position}
    \dot{x}(t) = -\frac{\partial V(x,\lambda)}{\partial x} + v(t) + \sqrt{2}\,\xi(t),
\end{equation}
where for simplicity we have set the friction $\gamma=1$, and energy scale $k_B T = 1$, such that the thermal diffusion $D = k_B T/\gamma=1$. \(\xi(t)\) is a zero-mean Gaussian white noise satisfying \(\langle \xi(t) \rangle = 0\) and \(\langle \xi(t)\xi(t') \rangle = \delta(t-t')\). The velocity $v(t)$ depends on the active model we choose; two generic choices we consider here are the Active Ornstein-Uhlenbeck process (AOUP) and the run-and-tumble process (RTP). For both cases, the velocity is a zero-mean coloured noise with covariance $\langle v(t)v(t')\rangle = \frac{D_v}{\tau} \exp{(-|t-t'|/\tau)}\,,$
where $\tau$ is the persistence time and $D_v$ is a self-propulsion ``diffusion'' constant. Up to second-order correlators, the statistics of the AOUP and RTP models coincide exactly. Higher moments, however, reveal that while the AOUP remains Gaussian, the RTP model becomes non-Gaussian~\cite{Put2019}. We define the P\'eclet number as $\mathrm{Pe} \equiv D_v / D$, a dimensionless parameter characterizing the strength of the active noise relative to the thermal noise. In the $\tau\to0$ limit the dynamics reduces to that of a passive particle.

\begin{figure}[t]
    \centering
    \includegraphics[width=\linewidth]{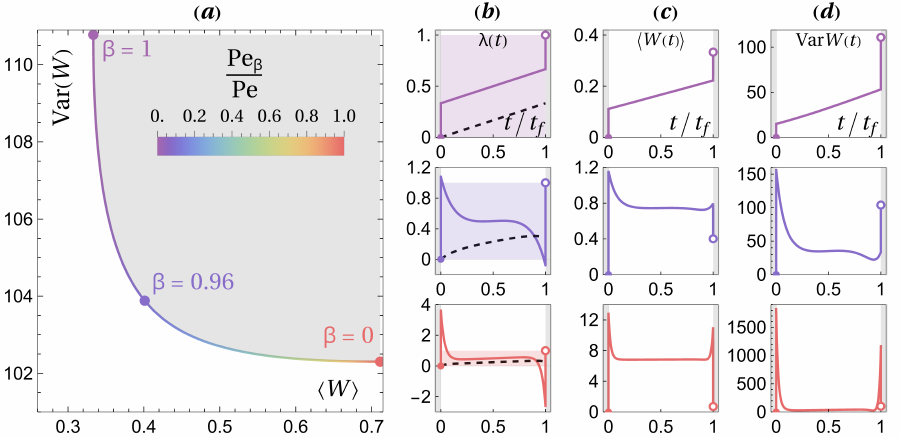}
    \caption{{\bf(a)} Work--fluctuation Pareto front coloured according to the effective P\'eclet number $\mathrm{Pe}_\beta$. {\bf(b)} Pareto-optimal protocols corresponding to three selected points on the front in {(a)}; coloured bands indicate the region $[0,\lambda_f]$, highlighting possible protocol overshoot. Dashed lines show the trapped particle's position $\langle x(t)\rangle$ {\bf(c,\,d)} Time-dependent average work and work fluctuations, respectively, corresponding to the optimal protocols in {(b)}. In panel columns (b-d), filled (open) points indicate the values of the corresponding quantities at $t=0$ ($t=t_f$). Parameters are $\lambda_f=1,\,t_f=1,\,\tau=0.5$, and $\mathrm{Pe}=200$. All quantities are expressed in dimensionless units obtained by rescaling time by $t_0=1/k$, energy by $k_B T=1$ and length by $\ell_0=\sqrt{k_B T/k}$ (see SM).}
    \label{fig:paretofig}
\end{figure}

We choose the control field as $\mathcal{U}(t) \equiv\lambda(t)$ and consider moving the trap centre from $\lambda_0 = 0$ to the target position $\lambda(t_f) = \lambda_f$ in a finite time $t_f$. Both at the start and end of the protocol, the system is in a nonequilibrium steady state (NESS). The ensemble-averaged position of the particle is then given by $\langle x(t)\rangle = k\int_{0}^{t}\lambda(t')\,e^{-k(t-t')}\mathrm{d}t'$. The average external input work a controller must supply during the protocol to overcome friction in the heat bath and potential energy changes of the particle is given by 
\begin{equation}
    \label{eq:work1}
    \langle W \rangle = \int_{\lambda_0}^{\lambda_f} \left\langle\frac{\partial V}{\partial \lambda}\right\rangle \,d\lambda = k \int_{0}^{t_f}\,\dot{\lambda}(t)(\lambda(t)-\langle x(t) \rangle)\,\mathrm{d}t\,,
\end{equation}
while the work fluctuations, characterised by the variance of the work input $ {\rm Var}{(W)}=\langle W^2 \rangle-\langle W \rangle^2$, is given by
\begin{equation}
    \label{eq:cov}
    \var{(W)} = k^2\int_{0}^{t_f}\int_{0}^{t_f}\dot{\lambda}(t')\,\dot{\lambda}(t)\,\covar(t,t')\,\mathrm{d}t'\mathrm{d}t\,.
\end{equation}
The positional covariance $\covar{(t,t')}=\langle x(t)x(t')\rangle$ can be computed exactly for the AOUP and RTP models~\cite{Loos2025,Loos2025_2}. To find the Pareto front, Eqns.~\eqref{eq:work1} and~\eqref{eq:cov} are combined into a single scalarised objective functional 
\begin{equation}
    \label{eq:combined_objective}
    \Omega[\lambda] = \beta \langle W\rangle\left[\lambda\right] + (1-\beta)\var{(W)}[\lambda]\,,
\end{equation}
where $\beta\in[0,1]$ denotes the relative weight of the work or its fluctuations in the optimisation, and where both objectives are made dimensionless by scaling by appropriate powers of $k_B T$. The constraints are encoded through $\mathcal{V}_1[\lambda] = \lambda(0)$ and $ \mathcal{V}_2[\lambda] = \lambda(t_f)-\lambda_f$. In the SM, we derive the optimal protocol $\lambda^*$ for any point on the front, parametrised by $\beta$, in dimensionless coordinates. We find that it generically admits the form $\lambda^*(t) = \mathcal{A}(\alpha) + \mathcal{B}(\alpha) t + \mathcal{C}(\alpha) \sinh{\left(\alpha (t-\frac{t_f}{2})\right)}$, where $\mathcal{A},\,\mathcal{B}$ and $\mathcal{C}$ are constants that depend on an inverse timescale $\alpha$, with $\alpha^2 = (1+\mathrm{Pe}_\beta)/\tau^2$. It is the characteristic inverse time over which the position fluctuations that matter for work decorrelate. Here, $\mathrm{Pe}_\beta = 2\mathrm{Pe}(1-\beta)/(2-\beta)$ is an effective activity that interpolates from the full activity $\mathrm{Pe}$ at $\beta=0$ to zero at $\beta=1$. Because $\beta$ enters $\lambda^*(t)$---and hence $\langle W\rangle$ and $\var{W}$--- only through $\alpha^{-1}$ two systems with identical $\tau$ but different bare $\mathrm{Pe}$ produce identical optimal protocols provided they sit at the $\beta$ values that equalise $\mathrm{Pe}_\beta$. The Pareto front thus reparametrises a family of physically distinct active systems onto one optimal-control curve. This front is shown in panel (a) of Fig.~\ref{fig:paretofig}, with corresponding optimal protocols shown in panel (b) for different points on the front. 

For $\beta=1$ (purple) the optimal protocol shows a constant velocity $\nu \equiv \dot{\lambda}(t) = \lambda_f/(2+t_f)$, with boundary jumps of size $\Delta\lambda$ at $t=0,\,t_f$, coinciding exactly with the optimal protocol for passive particles ($\mathrm{Pe}=0$)~\cite{Schmiedl2007,Loos2025}, where only the work is considered as a cost function; this situation therefore corresponds to an exceptional limiting case rather than a single generic behaviour.

Decreasing $\beta$ leads to protocols that overshoot the target $\lambda_f$ already at $t=0$, incurring a large initial work cost (Fig.~\ref{fig:paretofig}(c))~\cite{Kamp2026}. The trap then backtracks towards values below $\lambda_f$ while the particle moves towards $\lambda_f$. As a result, the instantaneous average power $\dot{W}$ is negative; energy stored during the initial jump is released back to the controller. Similarly, the initial jumps lead to a large increase in the work fluctuations, which are subsequently lowered during protocol relaxation. Particle trajectories that initially received more energy from the jump return energy faster to reduce fluctuations. When the trap centre again coincides with the mean position, the latter achieves its maximum and then reverses its motion. This shows that when the control goal allows active fluctuations to dominate system dynamics (i.e., high \emph{effective} $\mathrm{Pe}_\beta$), the optimal protocol shows a regenerative braking mechanism, where energy is recovered through an anticipatory mechanism; the controller `predicts' where the active particle moves to based on its persistence. In the fast driving limit, $\alpha t_f/2 \ll 1$, the critical Peclet number for regenerative braking to occur is ${\rm{Pe}}_\beta^{(c)} = (\tau + 1) (\tau+ \frac{t_f}{2})$, see SM.


Conversely, in the slow driving limit the average work and fluctuations scale asymptotically as $\langle W\rangle \sim \lambda_f^2/t_f$ and $\var(W) \sim 2\langle W\rangle(1+\mathrm{Pe})$, causing the Pareto front to contract to a single point and showing that the trade-off is purely a finite-time feature. The active contribution then enters instead through the effective diffusion coefficient $D_{\rm eff}=D+D_v$, which renormalises the transport geometry~\cite{Schmiedl2007,bb,addri}.

Interestingly, the optimal protocol away from the boundaries coincides in the limits $\tau \to 0$ and $\tau \to \infty$, where it reduces to the familiar linear ramp (purple curve in Fig.~\ref{fig:paretofig}(b)). In both regimes, activity plays a negligible role: for $\tau \to 0$ the dynamics reduce to those of a passive particle, while for $\tau \to \infty$ the motion is effectively ballistic on the protocol timescale, rendering active control unnecessary. In the high-activity regime, $\mathrm{Pe} \gg 1$, the protocol again becomes linear, but with a modified constant velocity, $\nu \sim \lambda_f/[2(1+\tau) + t_f]$, independent of the activity. The key distinction from the passive case, however, lies in the boundary behaviour: the boundary layers become sharply localized exponentials of width $\alpha^{-1}$, approaching Dirac delta contributions akin to optimal protocols in underdamped systems~\cite{Gomez2008}; the finite persistence time of the active force acts as a memory scale, so the optimal protocol develops the same boundary-layer structure as the inertial case. These singular boundary layers retain their characteristic over- and undershoots, with amplitudes scaling as $\Delta\lambda \sim \nu \sqrt{\mathrm{Pe}_\beta}$ for $\beta \neq 1$.

\emph{Quantum dot---}To illustrate our framework in a regime where energy discreteness is essential, we consider a quantum-dot heat engine with a single energy level coupled to a metallic lead. The lead temperature $T_\nu$ switches \emph{instantaneously}, between values $\nu\in\{c,h\}$, with $T_c < T_h$, on two intervals (strokes) $I_c=[0,t_f]$ and $I_h=[t_f,2t_f]$ (see Fig.~\ref{fig:cartoons}(b)). The control parameter $\epsilon(t)$ denotes the energy of the dot level measured relative to the (fixed) chemical potential of the lead, and can be experimentally tuned by an external gate voltage~\cite{Josefsson2018}. We impose cyclic operation, satisfying $p(0)=p(2t_f)$ and $\epsilon(0)=\epsilon(2t_f)=\epsilon_0$, and allow $\epsilon(t_f)=\epsilon_f$ to vary freely. Within each stroke, the occupation probability $p_\nu(t)$ evolves according to the master equation
\begin{equation}
\dot p_\nu+p_\nu=(1+e^{\epsilon(t) /T_\nu})^{-1}, \quad \nu \in \{c, h\},\label{mastereq}
\end{equation}
with time measured in tunneling units and $k_B=1$. Adopting the convention that work and heat are positive when entering the system and defining the cycle average $\langle f\rangle_\nu \equiv (2t_f)^{-1}\!\int_{I_\nu} f(t)\,\mathrm{d}t$, we compute the power $P=\sum_\nu \langle \dot\epsilon\,p_\nu\rangle_\nu$, heat flux $\dot Q_\nu=\langle \epsilon\,\dot p_\nu\rangle_\nu$ and entropy production rate $\dot\sigma=-\sum_\nu \frac{\dot Q_\nu}{T_\nu}$. Operation as a heat engine corresponds to $P < 0$ and $\dot Q_h > 0$; the engine efficiency is defined as $\eta = - P/\dot Q_h$, and is upper bounded by Carnot efficiency $\eta_C = 1 - T_c/T_h$.

We now set the control field as $\mathcal{U}(t) \equiv \epsilon(t)$ and consider two competing cost functionals: the average power $\mathcal{C}_1[\epsilon] \equiv P[\epsilon]$ and the dissipation rate $\mathcal{C}_2[\epsilon] \equiv T_c\,\dot\sigma[\epsilon]$, where the prefactor $T_c$ ensures both functionals share the same units. The Pareto front is then traced by the scalarised objective
\begin{equation}
\Omega[\epsilon] = \gamma\,P[\epsilon] + (1-\gamma)\,T_c\,\dot\sigma[\epsilon], \qquad \gamma \in [0,1]
\end{equation}

The single-objective limits $\gamma = 1$ and $\gamma = 0$ recover, respectively, the maximum-power and minimum-dissipation protocols. Two constraints close the optimisation: cyclic continuity enforces $\mathcal{V}_1[\epsilon] = p(2t_f) - p(0) = 0$, while equal-duration strokes are imposed by $\mathcal{V}_2[\epsilon] = 2t_f^{\nu} - t_f = 0$. Applying the optimality condition Eq.~(\ref{eq:functional_stationarity}) within each stroke yields a constant of motion along the optimal trajectory,
\begin{equation}
\frac{\dot p_\nu^{\,2}}{(1-p_\nu-\dot p_\nu)(p_\nu+\dot p_\nu)}=\frac{2 t_f\,\alpha_\nu}{\Lambda_\nu\,T_\nu} \equiv \omega_\nu^2(\gamma)\label{eq:constant_motion},
\end{equation}
\noindent where $\alpha_\nu$ relates to the dimensionless constant of motion and $\Lambda_c(\gamma) = 1$, $\Lambda_h(\gamma) = \gamma +(1-\gamma)\,T_c/T_h$. Inverting Eq.~\eqref{eq:constant_motion} gives $\dot p_\nu$ as a function of $p_\nu$, from which the optimal protocol $\epsilon^*(t)$ and its jumps $\Delta \epsilon$ follows through Eq.~(\ref{mastereq}) (see SM). Solving Eq.~(\ref{eq:functional_stationarity}) jointly with $\mathcal{V}_1, \mathcal{V}_2$ determines $(\alpha_c, \alpha_h, \epsilon_f)$ for each $\gamma$, parametrising the Pareto front and assigning an optimal protocol to each point. Importantly, at $\gamma=1$ both strokes follow the symmetric scaling $\omega_\nu \propto T_\nu^{-1/2}$, recovering the cyclic extension of the single-stroke optimum of Ref.~\cite{Esposito_2010}. While decreasing $\gamma$ progressively lowers the hot-stroke Pareto weight from 1 toward $1-\eta_C$, the cold stroke remains unchanged, thereby inducing a structural asymmetry between the two strokes.

Although the resulting protocol admits complicated closed-form solutions (see SM), its structure becomes more apparent in the regime where the dot energy is initialised at the lead's chemical potential ($\epsilon_0 = 0$) and a finite thermal gradient $T_h/T_c$ exists. In this regime, the Pareto-optimal protocol across the entire Pareto front is well-approximated by
\begin{equation}
    \label{cold}
    \epsilon_\nu^{*}(t;\gamma) \approx -T_\nu^{\text{eff}}(\omega_\nu,\gamma) \ln \Pi_\nu(t;\gamma), \qquad \nu\in\{h,c\}
\end{equation}
with effective temperatures $T_h^{\text{eff}} = T_h$ and $T_c^{\text{eff}}(\omega_\nu,\gamma) > T_c$. Here, $\Pi_\nu(t;\gamma)$ are second-order polynomials in time whose coefficients depend only on $\omega_\nu(\gamma)$. The deviation $T_c^{\mathrm{eff}} > T_c$ originates from the nonlinearity of the Fermi mapping, increasingly pronounced as the cold stroke approaches the band edge, whose higher-order corrections are absorbed into $T_c^{\mathrm{eff}}$. Notably, the entire $\gamma$-dependence is contained only in $\omega_\nu(\gamma)$, mirroring the role of the inverse timescale $\alpha^{-1}$ in the active-particle system. This defines once again a family of equivalence classes of optimal control problems: systems with different physical parameters but matching $\omega_\nu(\gamma)$ produce identical optimal protocols throughout the cycle. 

\begin{figure}[t]
    \centering
    \includegraphics[width=1\linewidth]{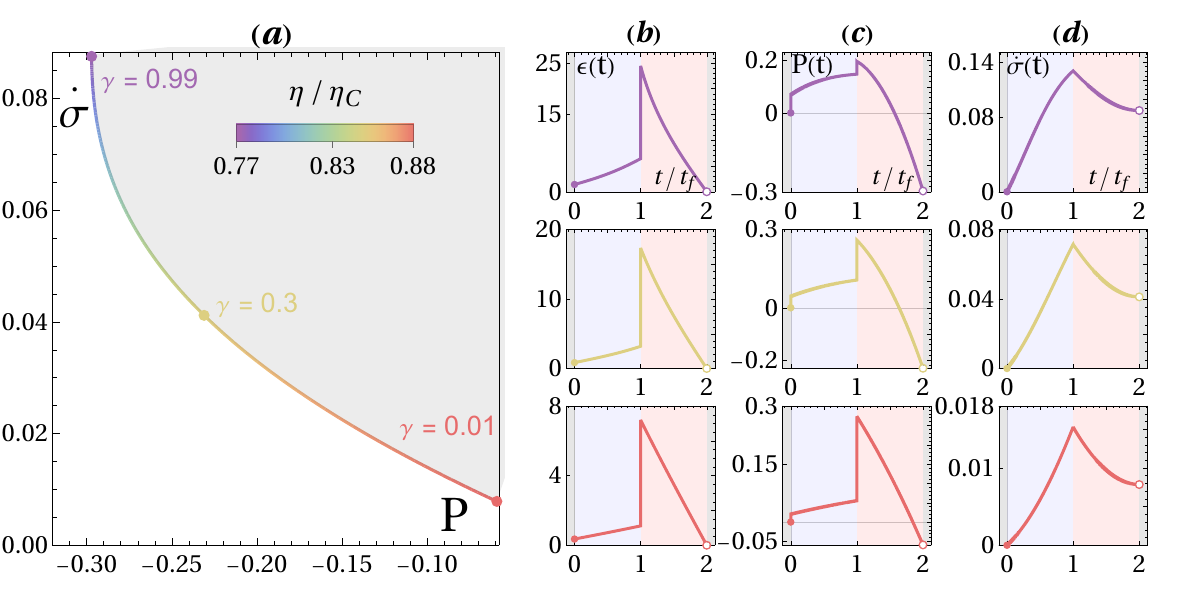}
    \caption{{\bf (a)} Power--dissipation Pareto front coloured according to the efficiency $\eta/\eta_C$. {\bf (b)} Pareto-optimal protocols corresponding to three selected points on the front in (a); shaded backgrounds indicate contact with the cold (blue) or hot (red) heat baths. {\bf(c,d)} Time-dependent power and entropy production, respectively, corresponding to the protocols in (b). Parameters are $T_h = 10,\,T_c = 1$, $t_f = 4$.}
    \label{fig:QDfig}
\end{figure}

The front and the corresponding efficiency are shown in panel (a) of Fig.~\ref{fig:QDfig}, while panel columns (b)--(d) display the optimal protocols $\epsilon^*(t)$, instantaneous power, and instantaneous entropy production along the front for different points. The optimization yields a universal cyclic structure: during the cold stroke, the protocol prepares the system for the subsequent work extraction by driving power and entropy production to grow monotonically, peaking at \(t=t_f\), where the system reaches its maximally displaced configuration. The hot stroke subsequently extracts work as the protocol drives the system back to the initial state. For small \(\gamma\), the protocol remains nearly linear, keeping the trajectory close to instantaneous equilibrium and thereby suppressing \(\dot{\sigma}\) while pushing \(\eta\) toward \(\eta_C\), in line with the low-dissipation geodesic structure of the thermodynamic manifold~\cite{tan_PRX_2023, sivak2023}. As \(\gamma\) grows, log-polynomial curvature becomes pronounced, accelerating energy build-up and generating sharper drops in \(P(t)\) that mark the maximum-power regime, at the cost of larger \(\dot{\sigma}\) and reduced \(\eta\). Importantly, along the front, $\omega_\nu^2$ and $\eta$ are in one-to-one correspondence, so the trade-off is governed by a single scale, as in the active particle case.

The jumps also grow monotonically with $\gamma$ since larger discontinuities store more out-of-equilibrium potential that is subsequently converted into power during the smooth evolution at the cost of additional dissipation. Crucially, over a complete cycle, these discontinuities act as thermodynamically free repositioning steps between strokes, storing and recovering internal energy reversibly to prepare the most favorable initial conditions without adding net dissipation.


Finally, the cycle duration $2t_f$ shapes the Pareto front through the effective parameter $\omega_\nu(\gamma)$. In the slow-driving limit $t_f \gg 1$, the protocol enters the low-dissipation regime: the entropy production scales as $\sigma \propto 1/t_f$, recovering the characteristic behavior of optimal finite-time control \cite{bb,sivak2023, tan_PRX_2023}. 
Taking additionally $T_h \to \infty$ contracts the front to a single point, indicating that the trade-off disappears in this limit, where the efficiency approaches $\eta_C$ and the extracted work approaches the reversible bound while the power vanishes. The opposite, fast-driving limit $t_f \to 0$, is bounded by the finite tunneling timescale of the dot, below which the cycle becomes unrealisable. Between these extremes, the extracted power is non-monotonic in $t_f$, attaining a maximum at $t_f^*$.

\emph{Conclusions---}We developed a general framework for multi-objective thermodynamic control by formulating optimal trade-offs between competing quantities as a variational problem for an effective cost functional. Rather than selecting a single optimum, our framework determines the full Pareto front and reveals a common structural feature: the relative weight assigned to each objective is absorbed into the physical parameters themselves, which combine into a single intrinsic scale governing the trade-off. This scale defines equivalence classes of control problems, such that systems with different parameters but the same scale share identical optimal protocols, suggesting a substantially more compact classification of nonequilibrium optimal-control problems. Importantly, the classical single-objective optima familiar from the literature emerge as the degenerate endpoints of the Pareto front~\cite{Schmiedl2007,Loos2025}, recovered as limiting cases of a broader variational structure.

This structure appears in two intrinsically nonequilibrium systems. For the active particle system, the Péclet number and persistence time combine into an effective timescale, and the optimal protocol develops anticipatory overshoots that exploit active-force memory as a regenerative-braking mechanism. For the quantum-dot engine, the bath temperatures and Pareto weights collapse into an effective parameter $\omega_\nu(\gamma)$ governing the full cycle, with inter-stroke jumps acting as thermodynamically free repositioning steps. In both cases, the full Pareto front is captured by a single function interpolating between competing strategies, with the usual single-objective optima arising only in limiting cases. Our predictions are within reach of existing experimental active-colloid~\cite{Buttinoni_2022,Goerlich_2025} and quantum-dot platforms~\cite{Josefsson2018}.

\section{Acknowledgments}
G.A.L.F and C.E.F. acknowledge the financial support from FAPESP under grants 2022/15453-0, 2022/16192-5, 2024/03763-0
2023/17704-2 and 2024/08157-0. The financial support from CNPq is also acknowledged. J.B. is supported by the Novo Nordisk Foundation with grant No. NNF18SA0035142.

\appendix

\bibliography{biblio}

\end{document}


\title{Supplemental Material for: Pareto-optimal control strategies in intrinsically nonequilibrium systems}

\author{Gustavo A. L. For\~ao}
\email[]{gforao@usp.br}
\affiliation{Universidade de São Paulo,
Instituto de Física,
Rua do Matão, 1371, 05508-090
São Paulo, SP, Brazil}

\author{Carlos E. Fiore}
\affiliation{Universidade de São Paulo,
Instituto de Física,
Rua do Matão, 1371, 05508-090
São Paulo, SP, Brazil}

\author{Jonas Berx}
\affiliation{Niels Bohr International Academy, Niels Bohr Institute,
University of Copenhagen, Blegdamsvej 17, 2100 Copenhagen, Denmark}

\date{\today}

\maketitle

\tableofcontents

\section{Pareto-optimal variational framework}\label{genframe}

We consider $N$ thermodynamic cost functionals
$\{\mathcal{C}_i[\boldsymbol{\mathcal{U}}]\}_{i=1}^N$ of the control field
$\boldsymbol{\mathcal{U}}(t)$, together with constraints
$\{\mathcal{V}_j[\boldsymbol{\mathcal{U}}]=0\}_{j=1}^{M_c}$ encoding fixed
endpoints or cyclicity of the control. Pareto-optimal protocols are stationary points of
\begin{equation}
\tilde\Omega[\boldsymbol{\mathcal{U}}]
=\sum_{i=1}^N \beta_i\,\mathcal{C}_i[\boldsymbol{\mathcal{U}}]
-\sum_{j=1}^{M_c} \mu_j\,\mathcal{V}_j[\boldsymbol{\mathcal{U}}],
\label{eq:em_scalarized}
\end{equation}
with Pareto weights $\beta_i\geq 0$, $\sum_i\beta_i=1$, and Lagrange
multipliers $\mu_j$; the condition $\delta\tilde\Omega=0$ yields Eq.~(\ref{eq:functional_stationarity})
of the main text,
\begin{equation}
\sum_i \beta_i\,\frac{\delta \mathcal{C}_i}{\delta \boldsymbol{\mathcal{U}}(t)}
= \sum_j \mu_j\,\frac{\delta \mathcal{V}_j}{\delta \boldsymbol{\mathcal{U}}(t)}.
\label{eq:em_stat}
\end{equation}

The key structural observation is the following. The multipliers $\mu_j$ are
not free: they are fixed, together with the protocol, by the constraints.
Multiplying all cost terms in Eq.~\eqref{eq:em_stat} by a constant factor $s>0$ leaves the same protocol optimal, with multipliers $s\mu_j$. The optimal protocol therefore depends only on \emph{ratios} of cost coefficients,
never on their overall scale. We now show that for the costs arising in
stochastic thermodynamics this observation collapses the full
parameter dependence onto a single effective scalar.

In Langevin or master-equation systems, all thermodynamic costs are built
from the same dynamical response to the control: moments of the
trajectory, or of the occupation probability. Different costs differ only
in how they weight this common structure. We therefore write
\begin{equation}
\mathcal{C}_i[\boldsymbol{\mathcal{U}};\boldsymbol\theta]
= \sum_{k=1}^{M} a_{ik}(\boldsymbol\theta)\,
\mathcal{S}_k[\boldsymbol{\mathcal{U}}],
\label{eq:em_modal_cost}
\end{equation}
with building blocks $\mathcal{S}_k$ fixed by the dynamics and
coefficients $a_{ik}$ carrying the dependence on the physical parameters
$\boldsymbol\theta$. The $\mathcal{S}_k$ may retain a subset of the
parameters (e.g. relaxation rates, protocol duration) so the reduction below
operates at fixed rates and duration; the Pareto weights enter only
through the coefficients, so sweeping the front is precisely where the
reduction is strongest. The scalarised objective inherits the form
$\Omega=\sum_k c_k\,\mathcal{S}_k$ with
$c_k(\boldsymbol\theta,\boldsymbol\beta)=\sum_i\beta_i a_{ik}(\boldsymbol\theta)$,
and by the scaling argument the optimal protocol can depend on
$(\boldsymbol\theta,\boldsymbol\beta)$ only through the $M-1$ ratios of
the coefficients $c_k$. We highlight that this same parametric reduction governs the global structure of the Pareto front: as the Pareto weights $\boldsymbol{\beta}$ are varied across the simplex, the optimal protocol traces the entire front through a single functional form, with only the coefficient ratios deforming continuously along it. The full Pareto front of a given system is thus generated by the same family of protocols sharing a fixed structure, parametrized by the intrinsic scalar(s) encoded in the ratios of the coefficients $c_k$. We now exemplify this result
using the two families realized in the main text. 

Costs quadratic in the driving velocity
$\nu(t)\equiv\dot{\boldsymbol{\mathcal{U}}}(t)$ typically take the form
\begin{equation}
\mathcal{S}_k = \int_0^{t_f}\!\!\int_0^{t_f} \nu(t)\,
e^{-\gamma_k|t-s|}\,\nu(s)\,dt\,ds,
\label{eq:em_bilinear_S}
\end{equation}
where each rate $\gamma_k$ is a relaxation mode of the dynamics: the
velocity at time $t$ is penalized through the memory it leaves in mode
$k$. With the endpoint constraint $\int_0^{t_f}\nu\,dt
=\Delta\boldsymbol{\mathcal{U}}$ and multiplier $\mu$, stationarity reads
\begin{equation}
2\sum_{k=1}^{M} c_k \int_0^{t_f} e^{-\gamma_k|t-s|}\,\nu(s)\,ds = \mu,
\qquad t\in(0,t_f),
\label{eq:em_fredholm}
\end{equation}
i.e.\ the weighted memory integral of the optimal velocity must be flat.
Since the kernels smooth whatever they act on, a bounded $\nu$ produces a
left-hand side that sags near the endpoints, where part of the memory is
cut off; flatness up to the boundary thus forces singular contributions at
$t=0,\,t_f$ (the protocol jumps well-known in optimal driving \cite{Schmiedl2007})
emerging here as the price of inverting a smoothing kernel on a finite
interval. This motivates the ansatz
\begin{equation}
\nu(t) = \sum_{r=1}^{M-1}\bigl[A_r e^{\omega_r t} + B_r e^{-\omega_r t}\bigr]
+ C + a\,\delta(t) + b\,\delta(t-t_f),
\label{eq:em_ansatz}
\end{equation}
\noindent where $a$ and $b$ are the magnitude of the jumps. Inserting \eqref{eq:em_ansatz} into \eqref{eq:em_fredholm} and using the elementary integral
\begin{equation}
\int_0^{t_f} e^{-\gamma|t-s|}e^{\omega s}\,ds
=
\frac{2\gamma}{\gamma^2-\omega^2}e^{\omega t}
-\frac{e^{-\gamma t}}{\gamma+\omega}
+\frac{e^{\omega t_f}}{\omega-\gamma}e^{-\gamma(t_f-t)},
\end{equation}
each trial term splits into two contributions: a bulk part proportional to the same function $e^{\omega t}$ that was inserted, and two boundary terms $e^{-\gamma_k t}$ and $e^{-\gamma_k(t_f-t)}$ that decay with the kernel rate $\gamma_k$ away from each endpoint. The latter arise only because the interval is finite. Matching Eq.~\eqref{eq:em_fredholm} then requires the bulk and boundary contributions to vanish separately.
Equation \eqref{eq:em_fredholm} holds if and only if
(i) the bulk exponentials cancel, i.e., 
\begin{equation}
R(\omega_r^2)\equiv\sum_{k=1}^{M}
\frac{2\,c_k\,\gamma_k}{\gamma_k^2-\omega_r^2}=0,
\label{eq:em_secular}
\end{equation}
which determines the $M-1$ admissible frequencies as the roots of a
rational function with poles at the $\gamma_k^2$;
(ii) the bulk constant matches the multiplier,
$C\sum_k 2c_k/\gamma_k=\mu/2$, fixing $\mu$; and
(iii) the boundary layers cancel mode by mode. The layer $e^{-\gamma_k t}$ is fed only by the kernel of mode $k$: the kernel of mode $j\neq k$ produces boundary layers $e^{-\gamma_j t}$ with $\gamma_j\neq\gamma_k$, which are linearly independent of $e^{-\gamma_k t}$ and therefore contribute nothing to its coefficient. The coefficient of $e^{-\gamma_k t}$ thus carries $c_k$ as an overall factor
\begin{equation}
c_k\Bigl[\,a-\tfrac{C}{\gamma_k}
-\textstyle\sum_r\bigl(\tfrac{A_r}{\gamma_k+\omega_r}
+\tfrac{B_r}{\gamma_k-\omega_r}\bigr)\Bigr]=0,
\label{eq:em_boundary}
\end{equation}
and similarly at $t_f$: for $c_k\neq0$ the brackets vanish, so the
amplitudes are fixed by $(\gamma_k,t_f,\omega_r)$ and the boundary data
alone. The counting closes ($2M$ boundary conditions, the constant
matching, and the endpoint constraint for the $2M+2$ unknowns), and the
conclusion follows: \emph{the optimal protocol depends on all physical
parameters and Pareto weights only through the roots of
Eq.~\eqref{eq:em_secular}}. For $M=2$,
\begin{equation}
\omega^2 = \gamma_1\gamma_2\,
\frac{c_1\gamma_2 + c_2\gamma_1}{c_1\gamma_1 + c_2\gamma_2},
\label{eq:em_omega_M2}
\end{equation}
a single scalar labelling the equivalence classes and parametrising the
position along the Pareto front. 

The active-particle problem provides a simple illustration of the
general framework. In this case the kernel contains two exponential
modes, corresponding to trap relaxation and active persistence,
with rates $\gamma_1=1$ and $\gamma_2=\tau^{-1}$, so that $M=2$ and
Eq.~\eqref{eq:em_omega_M2} applies directly. Identifying the kernel
weights with the coefficients $c_1=A$ and $c_2=B$ of
Eq.~\eqref{eq:combined_kernel}, the combinations
$c_1\gamma_1+c_2\gamma_2$ and $c_1\gamma_2+c_2\gamma_1$ simplify to
\begin{equation}
c_1\gamma_1+c_2\gamma_2=\frac{2-\beta}{2},
\qquad
c_1\gamma_2+c_2\gamma_1=
\frac{2-\beta}{2\tau}\left(1+\mathrm{Pe}_\beta\right),
\end{equation}
where
\begin{equation}\label{peclet1}
\mathrm{Pe}_\beta=
\frac{2(1-\beta)\mathrm{Pe}}{2-\beta}.
\end{equation}
Substituting these expressions into Eq.~\eqref{eq:em_omega_M2} yields
\begin{equation}
\omega^2=
\frac{1+\mathrm{Pe}_\beta}{\tau^2}
\equiv \alpha^2,
\end{equation}
recovering exactly the $\alpha^2$ stated in the main text. The general framework thus fixes the intrinsic scalar $\alpha^2$ and guarantees that the protocol depends on the parameters only through it; the explicit inversion yielding the closed-form optimal protocol $\lambda^*(t)$ is carried out in Sec.~\ref{subsec:optimal_protocol_Pareto}.

The location of the root encodes the character of the protocol. When all
$c_k>0$, $R$ is increasing between consecutive poles
[$dR/d\omega^2=\sum_k 2c_k\gamma_k/(\gamma_k^2-\omega^2)^2>0$], so exactly
one root lies in each interval $(\gamma_k^2,\gamma_{k+1}^2)$ and moves
monotonically with the coefficient ratios. If one cost enters with
negative coefficient, the root can exit this interval through the larger
pole while remaining a monotonic function of the ratio: for $M=2$,
Eq.~\eqref{eq:em_omega_M2} is a monotonic M\"obius function of $c_1/c_2$
that exceeds $\max(\gamma_1^2,\gamma_2^2)$ precisely when one coefficient
is negative. The active particle realises this second case (see
Sec.~\ref{subsec:optimal_protocol_Pareto}): there $\alpha^2=\omega^2$ grows from $\tau^{-2}$ up to $(1+\mathrm{Pe})\tau^{-2}$ along the front, and the
boundary layers of width $\alpha^{-1}$ sharpen accordingly, approaching
the singular structure of underdamped optimal control. When $c_2\to0$ the
root collides with the pole $\gamma_2^2$, the corresponding bulk mode
drops out, and the constant-velocity solution of
Ref.~\cite{Schmiedl2007} is recovered. Physically, the position of the root relative to the poles determines the interior structure of the protocol: a root lying between the poles yields a relaxation profile interpolating between the two modes, whereas a root located beyond the largest pole (as in the active-particle case) drives the boundary layers of width $\omega^{-1}$ toward delta-like spikes, giving rise to the sharp boundary overshoots shown in Fig.~\ref{fig:paretofig} in the main text. If the root were to cross zero, the smooth interior would become oscillatory, signalling a qualitative transition that does not occur in the present examples but remains accessible within the general framework.

The second family used to exemplify the framework consists of local densities,
$\mathcal{S}_k=\int_0^{t_f}F_k(\boldsymbol{\mathcal{U}},
\dot{\boldsymbol{\mathcal{U}}})\,dt$, for which stationarity gives the
Euler--Lagrange equation of
$\mathcal{L}_{\rm eff}=\sum_k c_k F_k$. Since $\mathcal{L}_{\rm eff}$ has
no explicit time dependence, each stroke carries a first integral
(Beltrami identity):
$\dot{\boldsymbol{\mathcal{U}}}\cdot
\partial_{\dot{\boldsymbol{\mathcal{U}}}}\mathcal{L}_{\rm eff}
-\mathcal{L}_{\rm eff}=K$. When a single block governs a stroke, $\mathcal{L}_{\rm eff}=c_1F_1$, the coefficient $c_1$ appears as an overall factor and therefore drops out of the Euler--Lagrange equation. As a result, the optimal shape within an isolated stroke is independent of the system parameters. Parameter dependence re-enters exclusively through the conditions that couple the strokes together (cycle closure, time
allocation), enforced by multipliers shared across strokes, which fix the
Beltrami constant $K$ of each stroke. The stroke trajectory thus obeys
\begin{equation}
\dot{\boldsymbol{\mathcal{U}}}\cdot
\partial_{\dot{\boldsymbol{\mathcal{U}}}}F_1-F_1
=\frac{K}{c_1}\equiv\omega^2:
\label{eq:em_local_scalar}
\end{equation}
the Beltrami constant measured in units of the cost coefficient is the
intrinsic scalar of the local family, playing exactly the role of the
root of Eq.~\eqref{eq:em_secular} in the bilinear family. The quantum-dot
engine realizes this case, one scalar $\omega_\nu^2$ per stroke, fixed by
the cyclicity and time-allocation conditions
(see Sec.~\ref{qdder}).

Two limitations should be stated. First, the weighted-sum scalarisation
\eqref{eq:em_scalarized} recovers the full Pareto front only where the
front is convex. For the bilinear family this holds whenever each
individual cost is a positive-semidefinite form of $\nu$---as is the case for work and
work variance---even if some modal coefficients $c_k$ are negative; for
general local densities convexity is not guaranteed, and non-convex
portions of the front, if present, are not reached by sweeping
$\boldsymbol\beta$~\cite{Solon2018}. Second, writing the costs as explicit functionals of the control presupposes a dynamics that can be integrated out. For the quantum-dot engine, this is achieved exactly by inverting the master equation to express the control in terms of the state probabilities, rendering the problem strictly local. For systems with genuinely nonlinear dynamics where such inversion is unavailable, the scaling argument still holds via Pontryagin's maximum principle. The modal decomposition can be performed directly on the control Hamiltonian, $H=\sum_k c_k\, h_k(\boldsymbol{x},\boldsymbol{\lambda}, \boldsymbol{\mathcal{U}})$, where $\boldsymbol{x}$ represents the system's state variables and $\boldsymbol{\lambda}$ denotes the associated costate variables. Under a global rescaling of the cost coefficients $c_k \to s c_k$ ($s>0$), the Pontryagin necessary conditions for optimality ($\partial H/\partial\boldsymbol{\mathcal{U}}=0$, $\dot{\boldsymbol{\lambda}} = -\partial H/\partial\boldsymbol{x}$) remain invariant provided the costates and multipliers rescale as $\boldsymbol{\lambda} \to s\boldsymbol{\lambda}$. The optimal protocol $\boldsymbol{\mathcal{U}}^*(t)$ is therefore invariant under a global rescaling of the cost coefficients, depending only on their relative ratios, so that only the geometry of the cost space (rather than its absolute scale) determines the structure of optimal control. Consequently, equivalence classes labelled by these ratios persist universally beyond the exactly solvable setting.

\section{Pareto-optimal protocol for active particles}\label{subsec:optimal_protocol_Pareto}

Section \ref{genframe} established that this problem realises the bilinear family with $M=2$ and that the optimal protocol is governed by the single scalar $\alpha^2$ of Eq.~(\ref{peclet1}). Here we carry out the explicit inversion that yields $\lambda^*(t)$ in closed form. Let us first make the problem dimensionless by rescaling all times and lengths by the characteristic trap relaxation time $t_0 = 1/k$ and size $\ell_0 = \sqrt{k_B T/k}$, respectively, i.e., $t' = t/t_0$ and $x' = x/\ell_0$. As a result, the protocol also rescales as $\lambda' = \lambda/\ell_0$, and $\tau' = \tau/t_0 = k\tau$ becomes dimensionless. The diffusion coefficient $D' = D t_0 / \ell_0^2 = D/k_B T$ becomes equal to one in these units, i.e., $D' = 1$, since we chose the friction coefficient to be equal to one, as well as $k_B T=1$. Finally, we introduce the dimensionless P\'eclet number $\mathrm{Pe} \equiv \omega^2 \tau /D = D_v/D$.

To find the Pareto-optimal protocols, we consider the average input work, $\langle W \rangle / (k_B T)$, and the work variance, $\mathrm{Var}(W) / (k_B T)^2$. Both quantities are rendered dimensionless by rescaling by the appropriate powers of $k_B T$. Doing so, they can be written as:
\begin{align}
    \frac{\langle W \rangle}{k_B T} &= \frac{1}{2} \int_0^{T} \int_0^{T} \dot{\lambda'}(t')\dot{\lambda'}(s') e^{-|t'-s'|}\mathrm{d}t' \mathrm{d}s'\,,\label{eq:work_dimensionless}\\
    \frac{\var{(W)}}{(k_B T)^2} &= \int_0^{T} \int_0^{T} \dot{\lambda'}(t')\dot{\lambda'}(s') \covar'{(t',s')}\mathrm{d}t' \mathrm{d}s'\,,\label{eq:var_dimensionless}
\end{align}
with $T = t_f/t_0$ and $\covar'{(W)}(t',s') = (k/k_B T) \covar{(t',s')}$. To find the Pareto front, we combine both observables into a single affine sum,
\begin{equation}
    \label{eq:combined_objective}
    \Omega[\lambda'] = \beta \frac{\langle W\rangle\left[\lambda'\right]}{k_B T} + (1-\beta)\frac{\var{(W)}[\lambda']}{(k_B T)^2}\,,
\end{equation}
with $\beta\in[0,1]$. Due to the similar double integration structure in Eqns.~\eqref{eq:work_dimensionless} and~\eqref{eq:var_dimensionless}, the combined objective can be simplified into the bilinear cost functional
\begin{equation}
    \label{eq:combined_objective_2}
    \Omega[\lambda] = \int_0^{T} \int_0^{T} \dot{\lambda}(t)\mathcal{K}_c(t,s)\dot{\lambda}(s)\mathrm{d}t\,\mathrm{d}s\,,
\end{equation}
which intrinsically depends on $\beta$. In the preceding equation, we have omitted the primes for notational simplicity and will continue to do so in all subsequent calculations. All non-primed quantities are therefore understood to be dimensionless, having been obtained through the appropriate rescaling. The combined kernel $\mathcal{K}_c(t,s)$ is given by 
\begin{equation}
    \label{eq:combined_kernel}
    \mathcal{K}_c(t,s) = A e^{-|t-s|} + B e^{-|t-s|/\tau}\,,
\end{equation}
with coefficients
\begin{equation}
    \begin{split}
        \label{eq:AB_coefficients}
        A &= \frac{2-\beta}{2} + \frac{(1-\beta) \mathrm{Pe}}{1-\tau^2}\,, \\
        B &= -\frac{\tau (1-\beta) \mathrm{Pe}}{1-\tau^2}\,.
    \end{split}
\end{equation}

The kernel $\mathcal{K}_c$ involves two distinct characteristic timescales, i.e., $M = 2$: the trap relaxation time $t_r = \gamma/k$---set to unity in our chosen units---and the persistence time $\tau$ associated with the active dynamics. The Pareto coefficient $\beta$ then tunes the contribution of both timescales, i.e., by setting $\beta=1$ the coefficient $B$ vanishes and the trap relaxation timescale is the only one present in the system. For $\beta = 0$ both relaxation channels contribute with maximal weight, so that the kernel retains the full interplay between the trap relaxation timescale and the persistence time. 

The Lagrangian used for the constrained optimisation is 
\begin{equation}
    \label{eq:functional_Pareto}
    \mathcal{L}[\lambda] = \int_0^{T} \int_0^{T} \dot{\lambda}(t)\mathcal{K}_c(t,s)\dot{\lambda}(s)\mathrm{d}t\,\mathrm{d}s - \mu \left(\int_0^{T}\dot{\lambda}(t)\mathrm{d}t - \lambda_f\right)\,,
\end{equation}
with $\mu$ a Lagrange multiplier. Setting the functional derivative $\delta\mathcal{L}/\delta\dot{\lambda}=0$, we find the following condition:
\begin{equation}
    \label{eq:fredholm}
    2 \int_0^{T} \dot{\lambda}(s) \mathcal{K}_c(t,s)\mathrm{d}s = \mu\,.
\end{equation}
The structure of Sec.~\ref{genframe} already fixes the solution to the form~\eqref{eq:diffeq_solution}; equivalently, and as an independent check, the integral equation~\eqref{eq:fredholm} can be reduced to a local ODE by annihilating its exponential kernels. We define the commuting linear operators $\hat{L}_1 = \left(\frac{\mathrm{d}^2}{\mathrm{d}t^2} - 1\right)$ and $\hat{L}_\tau = \left(\frac{\mathrm{d}^2}{\mathrm{d}t^2} - \tau^{-2}\right)$, with $\left[\hat{L}_1,\,\hat{L}_\tau\right] = 0$, where in general it holds that $\hat{L}_a e^{-a|t-t'|} = -2 a \delta(t-t')$ for any $a\in\mathbb{R}$. After applying both operators to Eq.~\eqref{eq:fredholm} and performing the resulting simplified integrations, we find the following differential equation for the protocol speed $\nu(t) = \dot{\lambda}(t)$:
\begin{equation}
    \label{eq:diffeq_speed}
    \left(A + \frac{B}{\tau}\right) \ddot{\nu}(t) - \left(\frac{A}{\tau^2} + \frac{B}{\tau}\right) \nu(t) = -\frac{\mu}{4\tau^2}\,,
\end{equation}
which can be written in the canonical form $\ddot{\nu}(t) - \alpha^2 \nu(t) = C$, where $C$ is a constant, $\alpha^2 = (1+\mathrm{Pe}_\beta)/\tau^2$, and the P\'eclet number is generalised as
\begin{equation}
    \label{eq:alpha_general}
    \mathrm{Pe}_\beta = \frac{2 (1-\beta) \mathrm{Pe}}{2-\beta}\,.
\end{equation}
The solution of the differential equation is then given by
\begin{equation}
    \label{eq:diffeq_solution}
    \nu(t) = \nu_c + c_1 e^{\alpha t} + c_2 e^{-\alpha t} + \Delta\lambda_0 \delta(t) + \Delta\lambda_f \delta(t-t_f)\,,
\end{equation}

\noindent We highlight that the Dirac-delta contributions are not introduced to satisfy the endpoint constraint, but rather they are structurally required to invert the smoothing kernel on the finite interval. As shown in Sec.~\ref{genframe}, the kernels map any bounded $\nu$ to a profile that sags at the endpoints, so reproducing the constant of Eq.~(\ref{eq:em_fredholm}) up to and including the boundary forces singular sources at $t=0$ and $t=t_f$. The endpoint and continuity data [Eqs.~(\ref{eq:cond1})--(\ref{eq:cond3})] then merely fix the amplitudes $\Delta\lambda_0$ and $\Delta\lambda_f$. This is distinct from the local (memoryless) family, where no kernel inversion occurs and the jumps arise directly from the boundary-supported $\delta V_j/\delta U$.

Due to the time reversal symmetry of the protocol~\cite{Loos2024}, it can be easily shown that the jumps must be of equal amplitude, i.e., $\Delta\lambda_0 = \Delta\lambda_f \equiv \Delta\lambda$, and that the coefficients $c_1$ and $c_2$ are related as $c_2 = c_1 e^{\alpha T}$.

We are thus left with three unknowns— the constant speed $\nu_c$, the jump amplitude $\Delta\lambda$ and the coefficient $c_1$---which require three independent conditions for their determination. A first condition can be determined by the finiteness of the protocol; it must hold that $\lambda_f = 2\Delta\lambda + \int_0^T \dot{\lambda}(t)\mathrm{d}t$. Plugging in the exact solution for $\nu(t)$ yields the condition
\begin{equation}
    \label{eq:cond1}
    \lambda_f = 2 \Delta\lambda + \nu_c T + \frac{2 c_1}{\alpha}\left(e^{\alpha T} -1\right)\,.
\end{equation}
The two other conditions can be found in the following manner. By inserting the solution~\eqref{eq:diffeq_solution} into Eq.~\eqref{eq:fredholm} and performing the resulting integrations, one can find that a particular structure of time-dependent exponentials emerges, only involving terms multiplying the factors $e^{-t}$ and $e^{-t/\tau}$ (and their time-reversed counterparts with identical coefficients). To guarantee that this sum of exponentials equals a single constant \emph{for all times}, the coefficients of each exponential involving a different timescale must be set identically to zero. Because $e^{-t}$ and $e^{-t/\tau}$ have different decay rates, there is no possible non-zero choice of coefficients that would allow them to sum to zero for all $t$. This yields the following two conditions: 
\begin{align}
    0& = \Delta\lambda - \nu_c - c_1 \left(\frac{1}{1+\alpha} + \frac{e^{\alpha T}}{1-\alpha}\right)\,, \label{eq:cond2}\\ 
    0& = \Delta\lambda - \tau\nu_c - \tau c_1 \left(\frac{1}{1+\alpha\tau} + \frac{e^{\alpha T}}{1-\alpha\tau}\right)\,. \label{eq:cond3}
\end{align}
Equations~\eqref{eq:cond1},~\eqref{eq:cond2} and ~\eqref{eq:cond3} can be solved to determine the unknown constants:
\begin{align}
    \Delta\lambda &= \frac{\lambda_f \alpha^2 \tau Q_+}{2 \left(\alpha Q_- (\tau + \frac{T}{2}) + (\alpha^2 - 1) R\right)}\,, \\
    \nu_c &= \frac{\lambda_f \alpha Q_-}{2 \left(\alpha Q_- (\tau + \frac{T}{2}) + (\alpha^2 - 1) R\right)}\,, \\
    c_1 &= \frac{\lambda_f \alpha (1 - \alpha^2) (\alpha^2 \tau^2 -1)}{2 \left( \alpha Q_- (\tau + \frac{T}{2}) + (\alpha^2 - 1) R\right)}\,,
\end{align}
where 
\begin{equation}
    \begin{split}
        R &= e^{\alpha T} (1+\alpha \tau) - (1-\alpha\tau)\,, \\
        Q_\pm &= e^{\alpha T} (1+\alpha \tau) (\alpha+1) \pm (1-\alpha\tau)(\alpha-1)\,.
    \end{split}
\end{equation}
Inserting these expressions into the protocol speed~\eqref{eq:diffeq_solution} and integrating yields the exact protocol $\lambda(t)$, as a function of the Pareto parameter $\beta$, which enters only in the characteristic inverse timescale $\alpha$. 

\subsection{Emergence of regenerative braking}\label{subsec:regenerative_braking}

\begin{figure}[htp]
    \centering
    \includegraphics[width=0.5\linewidth]{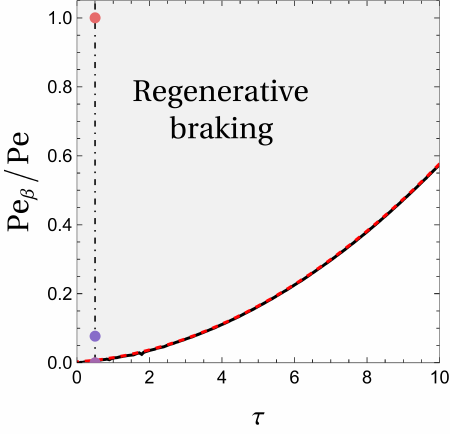}
    \caption{Emergence of regenerative braking (shaded region) based on the exact threshold~\eqref{eq:regenerative_braking_threshold} (black line) and the approximation~\eqref{eq:braking_approximation} (red, dashed line) for $t_f = 1$; they coincide almost identically. The vertical dot-dashed line at $\tau = 0.5$ shows that the Pareto front in the main text exhibits regenerative braking already for very small values of $\mathrm{Pe}_\beta$. The coloured points correspond to the three chosen parameter values of the representative protocols in the main text.}
    \label{fig:braking}
\end{figure}

The conditions for a protocol to exhibit regenerative braking are: \emph{(i)} the active particle must show a time interval in which it decelerates and \emph{(ii)} the instantaneous average power must be negative during this interval. Since the trap initially jumps to a position away from the particle at $t=0$ and starts backtracking at $t=0^+$ while the active particle is moving toward its goal, it immediately starts to decelerate. It then remains to check when the instantaneous average power becomes negative at $t=0^+$. From the definition of the power $\langle \dot{W} \rangle = k \dot{\lambda}(t)(\lambda(t)-\langle x(t) \rangle)$, we see that it is negative at $t=0^+$ when $\dot{\lambda}(0^+)<0$, since $\langle x(0^+)\rangle = 0$ and $\lambda(0^+) > 0$. This can be reduced to the implicit equation
\begin{equation}
    \label{eq:regenerative_braking_threshold}
    \tanh{\left(\frac{\alpha t_f}{2}\right)} = \frac{\alpha [(\alpha^2-1)\tau^2 - \tau -1]}{\tau+1}\,,
\end{equation}
which represents the phase boundary between the regenerative braking regime and the purely dissipative regime. In the limit where $\alpha t_f/2 \ll 1$ (fast driving), we can expand the l.h.s. of the equation and insert $\alpha^2 = (1+\mathrm{Pe}_\beta)/\tau^2$, which leads to the \emph{critical} effective P\'eclet threshold for regenerative braking to occur, i.e.,
\begin{equation}
    \label{eq:braking_approximation}
    \mathrm{Pe}_\beta^{(c)} = (\tau + 1) (\tau+ \frac{t_f}{2})\,.
\end{equation}
The condition~\eqref{eq:regenerative_braking_threshold} and its fast driving approximation~\eqref{eq:braking_approximation} are shown in Fig.~\ref{fig:braking}.
For $\mathrm{Pe}_\beta < \mathrm{Pe}^{c}_\beta$, the persistence time of the active drive is too weak to overcome the threshold set by the trap response and the protocol duration; the system thus remains in the trap-dominated regime, and the optimal protocol qualitatively resembles the canonical passive particle ($\mathrm{Pe}=0$) protocol~\cite{Schmiedl2007}. Conversely, for $\mathrm{Pe}_\beta > \mathrm{Pe}^{c}_\beta$, activity is strong enough (for that same persistence) to push the system into the opposite regime where the particle can overcome the trap response more effectively, and regenerative braking emerges.

\section{Pareto-optimal protocol for the quantum dot engine}\label{qdder} To find the Pareto-optimal protocols, we combine the average injected power $P[\epsilon]$ (such that minimizing $\Omega$ \emph{maximizes} the extracted work) and the entropy production rate into a single affine sum,
\begin{equation}
\Omega[\epsilon] = \gamma\,P[\epsilon] + (1-\gamma)\,T_c\,\dot\sigma[\epsilon], \label{eq:pareto_sum}
\end{equation}
with $\gamma \in [0,1]$. 
By expressing the mean power as $P = \sum_\nu (2t_f)^{-1}\int_{I_\nu} \dot\epsilon p_\nu\,dt$ and the entropy production as $T_c\,\dot\sigma = -\dot Q_c - (T_c/T_h)\dot Q_h$ (where $\dot Q_\nu = (2t_f)^{-1}\int_{I_\nu} \epsilon \dot p_\nu\,dt$), we apply integration by parts to the power term: $(2t_f)^{-1}\int_{I_\nu} \dot\epsilon p_\nu\,dt = (2t_f)^{-1}[\epsilon p_\nu]_{I_\nu} - (2t_f)^{-1}\int_{I_\nu} \epsilon \dot p_\nu\,dt$. 
Substituting these relations into Eq.~\eqref{eq:pareto_sum}, $\Omega$ decomposes into stroke contributions $\Omega = \Omega_c + \Omega_h$ of the form
\begin{equation}
\Omega_\nu[\epsilon] = (2t_f)^{-1}\gamma[\epsilon(t) p_\nu(t)]_{\partial I_\nu} - (2t_f)^{-1}\Lambda_\nu(\gamma)\int_{I_\nu}\epsilon\,\dot p_\nu\,dt, \label{eq:stroke_contrib}
\end{equation}
where the first term represent the scaled internal energy (per time) changes at the stroke limits, and the Pareto weights are defined as
\begin{equation}
\Lambda_c(\gamma) = 1, \qquad \Lambda_h(\gamma) = \gamma +(1-\gamma)\,T_c/T_h. \label{eq:pareto_weights}
\end{equation}
The coefficient $\gamma$ tunes the relative weight of the hot stroke: $\Lambda_h$ ranges from $\Lambda_h = T_c/T_h$ at $\gamma=0$ to $\Lambda_h = 1$ at $\gamma=1$, recovering the cyclic formulation of \cite{Esposito2010}.

Expressing the protocol $\epsilon(t)$ as function of $p_\nu$ and substituting into Eq.~\eqref{eq:stroke_contrib} defines the effective Lagrangian density
\begin{equation}
\mathcal{L}_\nu^{\mathrm{eff}}(p_\nu, \dot p_\nu; \gamma) = -\frac{2t_f}{\Lambda_\nu(\gamma)\,T_\nu}\,\dot p_\nu\ln\!\frac{1-p_\nu-\dot p_\nu}{p_\nu + \dot p_\nu},
\label{effqd}
\end{equation}
whose Beltrami identity yields the constant of motion $\omega_\nu^2 (\gamma)$ such that
\begin{equation}
\frac{\dot p_\nu^{\,2}}{(p_\nu+\dot p_\nu)(1-p_\nu-\dot p_\nu)} ={\frac{2 t_f \alpha_\nu}{\Lambda_\nu T_\nu}}\equiv \omega_\nu^2 (\gamma) , \label{eq:beltrami}
\end{equation}
where we define $\omega_\nu^2 (\gamma)$ to absorb the entire $\gamma$-dependence of each stroke. 
Solving Eq.~\eqref{eq:beltrami} for $\dot p_\nu$ and combining with the dynamic equation yields the optimal occupation
\begin{equation}
p_\nu^{\textrm{opt}}(t) = \frac{1 \pm e^{\epsilon(t)/(2T_\nu)}\sqrt{\omega_\nu^2 (\gamma)}}{1 + e^{\epsilon(t)/T_\nu}}, \label{eq:optimal_occupation}
\end{equation}
with the upper (lower) sign for the cold (hot) stroke. The upper sign applies to upward strokes ($\epsilon(t)$ increasing in the cold reservoir, $\epsilon_{t_f} > \epsilon_0$), corresponding to occupation above the instantaneous equilibrium; the lower sign applies to downward strokes (hot reservoir), with occupation below equilibrium.

Comparing Eq.~\eqref{eq:optimal_occupation} to the cyclic boundary condition forces discontinuous jumps in $\epsilon(t)$ at $t=0$ and $t = t_f$. Each physical macroscopic jump is composed of a quasi-static thermal contribution due to the bath quench, and kinetic contributions from the finite-time driving of both adjacent strokes. The total amplitude is given by

\begin{align}
    \Delta\epsilon(t^*) = \Delta\epsilon_\mathrm{qs} - 2T_h\,\mathrm{arcsinh}\!\left[\sqrt{\omega_h^2 (\gamma)}\cosh\!\left(\frac{\epsilon_h^*}{2T_h}\right)\right] - 2T_c\,\mathrm{arcsinh}\!\left[\sqrt{\omega_c^2 (\gamma)}\cosh\!\left(\frac{\epsilon_c^*}{2T_c}\right)\right], \label{eq:energy_jumps}
\end{align}
where $\Delta\epsilon_\mathrm{qs} = \epsilon_h^* - \epsilon_c^*$ is the required energy shift to maintain the instantaneous occupation invariant across the temperature transition, and $\epsilon_\nu^*$ denotes the instantaneous equilibrium energy corresponding to that occupation at temperature $T_\nu$. Between jumps, integrating $dt=dp/\dot p_\nu^\pm$ yields the relation
\begin{equation}
t = F_\nu^\pm[p_\nu(t); \omega_\nu^2 (\gamma)] - F_\nu^\pm[p_\nu^{\mathrm{ini}}; \omega_\nu^2 (\gamma)], \label{eq:implicit_time}
\end{equation}

\noindent where

\begin{align}
F_\nu^\pm(p; \omega_\nu^2) &= -\frac{1}{2}\ln[p(1-p)] \pm \frac{1}{\sqrt{\omega_\nu^2}}\arctan\!\left[\frac{1-2p}{\sqrt{\omega_\nu^2+4p(1-p)}}\right] \nonumber \\
&\quad \pm \frac{1}{2}\,\mathrm{arctanh}\!\left[\frac{\sqrt{\omega_\nu^2}(1-2p)}{\sqrt{\omega_\nu^2+4p(1-p)}}\right], \label{eq:f_function}
\end{align}

\noindent with the upper (lower) sign matching the cold (hot) stroke. The optimal protocol is then obtained by inverting Eq.~\eqref{eq:implicit_time} for $p_\nu(t)$ and substituting into

\begin{equation}
\epsilon_\nu^{\textrm{opt}}(t) = T_\nu\ln\!\left[\frac{1-p_\nu(t)-\dot p_\nu(t)}{p_\nu(t) + \dot p_\nu(t)}\right]. \label{eq:optimal_eps}
\end{equation}

\noindent The three unknowns $(\alpha_c, \alpha_h, \epsilon^*)$ are determined for each $\gamma$ by the duration constraints $t_f^c(\alpha_c, \epsilon_0, \epsilon^*) = t_f^h(\alpha_h, \epsilon_0, \epsilon^*) = t_f$ and the optimality condition $\partial\Omega/\partial\epsilon_{t_f} = 0$, parametrizing the full Pareto front.
\newline
\newline
\noindent \textbf{Origin of the log-polynomial form---} Notably, the first integral of motion Eq.~(\ref{eq:beltrami}) defines a closed curve in the ($p_\nu$, $\dot p_\nu$) plane parametrized by the single quantity $\omega_\nu(\gamma)$. Inverting the Fermi-Dirac master equation Eq.~(\ref{mastereq}) yields  the exact protocol in Eq.~(\ref{eq:optimal_eps}). In the experimentally-grounded regime explored in the main text, $X_\nu(t) \equiv p_\nu(t) + \dot p_\nu(t)$ varies smoothly along each stroke and admits a low-order polynomial representation, whose coefficients are controlled by $\omega_\nu(\gamma)$ and the stroke endpoints. Substituting this approximation, the protocol takes the compact form seen in Eqs.~(\ref{cold}), where the effective temperature $T_\nu^\textrm{eff}(\gamma)$ absorbs the subleading $\ln (1-X_\nu)$ contribution from the Fermi inversion into a single rescaled prefactor. Physically, the optimal protocol tries to minimize the dissipation by suppressing rapid departures from instantaneous equilibrium, as any excess thermodynamic velocity directly contributes to entropy production. The optimum therefore corresponds to the slowest occupation trajectory compatible with the prescribed cycle time, with the conserved Beltrami quantity $\omega_\nu^2$ setting its characteristic velocity scale. Since the control tracks the instantaneous-equilibrium occupation, $X_\nu=p_\nu+\dot{p}_\nu=p_{\rm eq}(\epsilon_\nu)$, which is a smooth and bounded Fermi function, the resulting trajectory is governed by a single relaxation scale and is accurately captured by a low-order polynomial. The apparent complexity of the protocol then arises solely from its representation in energy space, where the Fermi mapping becomes increasingly nonlinear near the band edges.

\bibliography{biblio}